# Long non-coding RNAs as a source of new peptides

Jorge Ruiz-Orera[1], Xavier Messeguer[2], Juan A Subirana[1,3], M.Mar Albà[1,4,*]

[1]Evolutionary Genomics Group, Research Programme on Biomedical Informatics (GRIB) - Hospital del Mar Research Institute (IMIM) - Universitat Pompeu Fabra (UPF), Barcelona, Spain. [2]Universitat Politècnica de Catalunya (UPC), Barcelona. [3]Real Academia de Ciències i Arts de Barcelona (RACAB). [4]Catalan Institution for Research and Advanced Studies (ICREA), Barcelona.

* To whom correspondence should be addressed.

Corresponding author contact information:
M.Mar Albà
ICREA Research Professor
Research Programme on Biomedical Informatics (GRIB) - Hospital del Mar Research Institute (IMIM),
Barcelona Biomedical Research Park (PRBB)
Barcelona 08003, Spain
Phone: +34933160516, FAX: +34933160550
Email: malba@imim.es

# ABSTRACT

Deep transcriptome sequencing has revealed the existence of many transcripts that lack long or conserved open reading frames and which have been termed long non-coding RNAs (lncRNAs). Despite the existence of several well-characterized lncRNAs that play roles in the regulation of gene expression, the vast majority of them do not yet have a known function. Motivated by the existence of ribosome profiling data for several species, we have tested the hypothesis that they may act as a repository for the synthesis of new peptides using data from human, mouse, zebrafish, fruit fly, *Arabidopsis* and yeast. The ribosome protection patterns are consistent with the presence of translated open reading frames (ORFs) in a very large number of lncRNAs. Most of the ribosome-protected ORFs are shorter than 100 amino acids and usually cover less than half the transcript. Ribosome density in these ORFs is high and contrasts sharply with the 3'UTR region, in which very often there is no detectable ribosome binding, similar to *bona fide* protein-coding genes. The coding potential of ribosome-protected ORFs, measured using hexamer frequencies, is significantly higher than that of randomly selected intronic ORFs and similar to that of evolutionary young coding sequences. Selective constraints in ribosome-protected ORFs from lncRNAs are lower than in typical protein-coding genes but again similar to young proteins. These results strongly suggest that lncRNAs play an important role in *de novo* protein evolution.

# INTRODUCTION

Studies performed over the past decade have unveiled a richer and more complex transcriptome than was previously appreciated (Okazaki *et al*, 2002; Carninci *et al*, 2005; Kapranov *et al*, 2007; Ponjavic *et al*, 2007). Thousands of long RNA molecules (>200 nucleotides) that do not display the typical properties of well-characterized protein-coding RNAs, and which have been named intergenic or long non-coding RNAs (lncRNAs), have been discovered in several eukaryotic genomes (Okazaki *et al*, 2002; Ponting *et al*, 2009; Cabili *et al*, 2011; Liu *et al*, 2012; Pauli *et al*, 2012; Ulitsky & Bartel, 2013). Intriguingly, with the exception of a few notable cases (Guttman & Rinn, 2012; Ulitsky & Bartel, 2013), the vast majority of lncRNAs do not have a known function.

A classical example of a lncRNA that has a well-characterized function is the X-inactive specific transcript (*Xist*), a regulator of X chromosome inactivation in eutherian mammals (Brockdorff *et al*, 1992). *Malat1*, another well-studied lncRNA, appears to be important for the organization of nuclear speckles (Tripathi *et al*, 2010). Other lncRNAs, such as *Hottip* or *Mistral*, have been proposed to have cis-regulatory roles, as their perturbation affects the expression of nearby genes. However, in general, the effect of lncRNAs on the expression of neighboring genes appears to be comparable to that observed between pairs of protein-coding genes (Cabili *et al*, 2011).

Interestingly, several independent studies have recently noted that a large fraction of lncRNAs associate with ribosomes (Ingolia *et al*, 2011; Juntawong *et al*, 2014; Bazzini *et al*, 2014; van Heesch *et al*, 2014). Deep sequencing of ribosome-protected fragments, or ribosome profiling, provides detailed information on the regions that are translated in a transcript (Ingolia, 2014). According to some studies, the patterns of ribosome protection indicate that lncRNAs are capable of translating short peptides (Ingolia *et al*, 2011; Bazzini *et al*, 2014; Juntawong *et al*, 2014) although others have reached different conclusions (Guttman *et al*, 2013). Perhaps the association with ribosomes is not so surprising when we consider that many lncRNAs have the same structure as classical mRNAs: they are transcribed by polymerase II, they are capped and polyadenylated, and they accumulate in the cytoplasm (van Heesch *et al*, 2014). Studies in mammals have revealed that, in contrast to the average protein-coding gene, most lncRNAs contain few introns, exhibit few sequence constraints, are expressed at low levels, tend to be tissue-specific, and they show limited phylogenetic conservation (Cabili *et al*, 2011; Derrien *et al*, 2012; Kutter *et al*, 2012; Necsulea *et al*, 2014).

The association of lncRNAs with ribosomes, and the fact that many of them appear to have arisen relatively recently in evolution, indicate that they could be an important source of new peptides. This hypothesis was initially proposed by Wilson and Masel after observing that ribosome profiling reads from a yeast experiment often mapped to intergenic transcripts (Wilson & Masel, 2011). Another study found evidence of translation of short species-specific ORFs located in non-genic regions (Carvunis *et al*, 2012). Xie and coworkers observed that several newly evolved protein-coding transcripts in humans were evolutionarily related to non-coding transcripts in macaque, establishing a link between non-coding RNAs and the evolution of new peptides (Xie *et al*, 2012). More generally, it is important to consider that *de novo* protein-coding gene evolution, which was once thought to be a very rare event, is now believed to be relatively common (Khalturin *et al*, 2009; Toll-Riera *et al*, 2009; Tautz & Domazet-Lošo, 2011; Long *et al*, 2013; Reinhardt *et al*, 2013). Recently emerged proteins tend to be very short and evolve under weak evolutionary constraints (Albà & Castresana, 2005; Levine *et al*, 2006; Cai *et al*, 2009; Liu *et al*, 2010; Xie *et al*, 2012; Palmieri *et al*, 2014), properties that we also expect to find in the putative ORF sequences of lncRNAs.

The idea that lncRNAs serve as a repository for the evolution of new peptides is appealing but the evidence is still fragmented. The ribosome profiling patterns of lncRNAs need to be carefully compared to those of protein-coding transcripts (codRNAs) to understand if lncRNAs are likely to be translated. In addition, the characteristics of the lncRNA ORF sequences should be compared to those observed in proteins that have emerged recently in evolution (lineage-specific proteins). Here we analyse ribosome profiling experiments performed in six different species and measure the sequence coding potential and selective constraints of the putatively translated ORFs in lncRNAs and codRNAs. The results strongly support a role for lncRNAs in the production of new peptides.

**RESULTS**

**Characterization of coding and long non-coding transcripts**

We obtained polyA+ RNA and ribosome profiling sequencing data from six different published experiments performed in diverse eukaryotic species, mouse (*Mus musculus*), human (*Homo sapiens*), zebrafish (*Danio rerio*), fruit fly (*Drosophila melanogaster*), *Arabidopsis* (*A. thaliana),* and yeast (*Saccharomyces cerevisiae*) (Table 1). After read mapping and transcript assembly we classified the expressed transcripts longer than 200 nucleotides into the following classes: 1. Coding RNAs (codRNAs), transcripts that mapped to Ensembl annotated protein-coding transcripts; 2. Annotated long non-coding RNAs (lncRNAs), transcripts that mapped to Ensembl annotated non-coding transcripts (only from genes not expressing or overlapping any coding transcript); 3. Novel lncRNAs, novel transcripts that did not overlap any annotated gene. We recovered all transcripts with expression level > 0.5 FPKM (Fragments Per Kilobase per total Million mapped reads). This guaranteed detection of ribosome association for the majority of coding transcripts (see next section) while yielding proportions of transcripts comparable to those reported in the original papers.

We detected hundreds of annotated lncRNAs in the vertebrate species (mouse, human and zebrafish) and a lower number of annotated lncRNAs (<100) in the other species (fruit fly, *Arabidopsis* and yeast) (Table 2). In addition, we identified a large number of novel lncRNAs, which amounted to 2,634 considering all species together. The inclusion of such lncRNAs resulted in a six-fold increase in the number of lncRNAs amenable for study in zebrafish and fruit fly and a two-fold increase in mouse. In yeast, we only found 3 annotated lncRNAs, but there were 69 novel ones. In the majority of the analyses we merged the annotated and the novel lncRNAs.

As expected, lncRNAs tended to be much shorter than codRNAs in all the species studied (Figure 1a, supplementary file 1 Table 1 and Figure 1). We also noted that both coding and non-coding transcripts tended to be longer in the three vertebrate species than in the other analyzed eukaryotic species. We found that most lncRNAs contain at least one short ORF (≥ 24 amino acids) and often several ORFs. The average ORF size in lncRNAs was between 43 and 68 amino acids depending on the species. Consistent with previous studies (Derrien *et al.*, 2012; Necsulea *et al.*, 2014) and with the exception of yeast, lncRNAs were expressed at significantly lower levels than codRNAs (Figure 1b, Wilcoxon rank-sum test, p-value <$10^{-5}$).

**Efficient detection of translation events by ribosome profiling**

Ribosome profiling, based on deep sequencing of ribosome-protected RNA sequences, has been shown to be a very powerful technique in the identification of translation events (Ingolia, 2014). Indeed, we found that the percentage of coding transcripts associated with ribosomes was >90 % in all species, with the highest values (>99%) in mouse and fruit fly (Table 2). In general, the results were similar for codRNAs encoding experimentally validated proteins (codRNAe) and for those encoding predicted proteins (codRNAne) (Supplementary File 1 Table 2). Pseudogenes had a lower rate of association with ribosomes than coding RNAs, but surprisingly, in species with many annotated pseudogenes, such as human, mouse and *Arabidopsis*, the majority of them showed association with ribosomes. This appeared to be a true signal; while pseudogenes will typically show sequence similarity to other functional copies in the genome, this result was based on uniquely mapped reads.

Ribosome profiling is based on deep sequencing and thus provides an unmatched level of resolution of the translated peptides when compared with current proteomics techniques. This is especially important for short proteins, which are difficult to detect by standard mass spectrometry methods (Slavoff *et al*, 2013). We used the ribosome-associated protein-coding RNA data to investigate the relationship between peptide detection by proteomics and protein length. We found that translated proteins between 24 and 80 amino acids long were systematically more difficult to detect than longer proteins in human and mouse proteomics databases (Table 3). This shows that short functional proteins are currently underrepresented in proteomics databases.

**Long non-coding RNA transcripts frequently associate with ribosomes**

The percentage of lncRNAs scanned by ribosomes (lncRNA_ribo) was surprisingly high in all the species studied (Table 2). For annotated lncRNAs, this percentage ranged from 40.4% in zebrafish to >90% in fruit fly and *Arabidopsis*. In general, the values were lower in novel lncRNAs, perhaps because these transcripts tend to be shorter than the annotated ones (Supplementary file 1 Table 1), and the detection of association with ribosomes is more difficult in short transcripts than in longer ones due to the lower number of ribosomes that can scan a short mRNA at a given time (Aspden et al., 2014; http://dx.doi.org/10.1101/002998). We found that the gene expression level can also be a limiting factor (Figure 2). Relatively few of the poorly expressed lncRNAs (<0.5 FPKM) in our datasets were associated with ribosomes (8.5% in mouse, 15.8% in human, and 6.4% in zebrafish). The percentage increased remarkably for highly expressed lncRNAs (>2 FPKM) (90.7% in mouse, 56.9% in human and 46.5% in zebrafish). A positive relationship between expression level and the capacity to detect ribosome association was also observed in the case of codRNAs (Figure 2).

As lncRNAs tend to be expressed at low levels and are short, we are probably underestimating their association with ribosomes. We compared the size and expression level of lncRNAs associated with ribosomes (lncRNA_ribo) and lncRNAs not associated with ribosomes (lncRNA_noribo). The former were in general longer and expressed at higher levels than the latter (Figure 1 and Supplementary file 1 Figure 1, lncRNA_ribo *versus* lncRNA_noribo), which is consistent with the existence of experimental limitations in the case of short and weakly expressed transcripts. In the most sensitive ribosome profiling experiments judging by the results obtained for codRNAs (mouse, fruit fly, *Arabidopsis*), the fraction of lncRNAs associated with ribosomes also happened to be the highest (Table 2).

The association with ribosomes affected the main lncRNA classes described in Ensembl v. 70 (Supplementary file 1 Table 3). For instance, the percentage of lincRNAs (long intervening non-coding RNAs) associated with ribosomes was 85.5% in mouse, 38.1% in human, and 33.3% in zebrafish, similar to the general figure for all lncRNAs taken together (81.9% in mouse, 43.4% in human, and 30.4% in zebrafish). The classes "processed transcripts" and "retained introns" displayed the highest ribosome association (i.e. 89.5% in mouse, 68.9% in human, and 42.2% in zebrafish), indicating that the majority of them are probably protein-coding. Many novel lncRNAs also showed evidence of being translated (one such example is displayed in Supplementary file 1 Figure 2A).

We collected a set of 29 human genes with non-coding functions described in the literature (Supplementary file 2 Table a) (Ponting *et al*, 2009; Ulitsky & Bartel, 2013; Fatica & Bozzoni, 2014). Many of these genes play roles in the regulation of gene expression in the nucleus and are thus unlikely to be translated. Only 5 of the genes were detected in the human polyA+ RNA-seq experiment (Hela cells): *Malat1*, *Pvt1*, *Neat1*, *Meg8*, and *Cyrano*. The first three genes mapped to ribosome profiling reads. In the case of *Malat1* the association was also observed in mouse and zebrafish (in the latter species *Malat1* was identified as a novel transcript), and in the case of *Pvt1* in mouse. In general, the small sample size precluded drawing any conclusions for this set.

Next we examined ribosome density in lncRNAs and codRNAs. To do this we calculated the mean translational efficiency (mean TE): the number of ribosome profiling reads in a transcript divided by RNA-seq transcript abundance (Ingolia *et al*, 2011). In order to account for variations in the relative size of the putatively translated regions in different transcripts (Supplementary file 1 Figure 3) we focused on regions covered by ribosome profiling reads. In mouse and human mean TE in lncRNAs and codRNAs was relatively similar (Figure 3). In fruit fly and yeast it was higher in codRNAs than in lncRNAs; and in zebrafish and *Arabidopsis* the opposite trend was observed (Wilcoxon rank-sum test, p-value <$10^{-5}$). The results were similar when we employed the maximum TE in 90 nucleotide windows or when we restricted the analysis to genes encoding a single transcript to avoid any possible biases due to reads mapping to different transcripts from the same gene (Supplementary file 1 Figure 4 and Table 4).

**LncRNAs show similar ribosome protection profiles to codRNAs**

The exact positions of ribosome profiling reads on the RNA can be used to delineate the regions that are being actively translated or to discover new functional ORFs (Chew *et al*, 2013; Guttman *et al*, 2013; Ingolia, 2014). Because the ribosome is released after encountering a stop codon, this technique can also be employed to identify novel C-terminal protein extensions (Dunn *et al*, 2013) or to evaluate if a predicted ORF is likely to correspond to a translated peptide (Guttman *et al*, 2013). In order to gain further insights into the translation patterns of lncRNAs, we compared the TE values in different transcript regions, including open reading frames (ORFs), putative 5' and 3' untranslated regions (UTRs), and the remaining parts of the transcript.

In order to obtain an unbiased picture, it was important to define the different regions in the same way in lncRNAs and codRNAs. In typical coding RNAs there is a main translated ORF that covers a large fraction of the transcript, sometimes accompanied by short upstream ORFs in the 5'UTR (Chew *et al*, 2013). However, lncRNAs may potentially encode several short peptides (Supplementary file 1 Table 1)(Ingolia *et al*, 2011). The minimum size of ORFs was set at 24 amino acids (75 nucleotides counting the STOP codon), as peptides of this size have been identified in genetic screen studies in humans (Hashimoto *et al*, 2001). We also considered both a primary ORF, defined as the ORF with the largest number of ribosome profiling reads, as well as any additional non-overlapping ORFs that mapped to ribosome profiling reads (rest of ORFs).

In codRNAs, the primary ORF showed a nearly perfect degree of agreement with the annotated protein; other metrics, such as using the ORF that exhibited the max TE, did not perform so well (Supplementary file 1 Figure 5). As expected, primary ORFs in lncRNAs were on average much shorter than primary ORFs in codRNAs (Supplementary file 1 Figure 6). Most of the putatively translated proteins in lncRNAs are less than 100 amino acids long. The number of lncRNAs with a primary ORF longer than 100 amino acids ranged between 8% and 22% in the vertebrate species and was even lower in the rest of species. Primary ORFs in lncRNAs also typically occupied a shorter fraction of the transcript than in codRNAs (Figure 4a).

Next we focused our attention on the differences between the primary ORF and the 5'UTR and 3'UTR regions. We defined the 3' untranslated region (3'UTR) as the sequence located immediately after the STOP codon of the primary ORF or the most downstream ORF associated with ribosomes. We used the same criteria to define the 5'UTR upstream from the initiation codon. In this analysis we included all transcripts containing at least one ORF associated with ribosomes (the primary ORF) and sufficiently long UTR regions as to detect ribosome profiling reads (>30 nucleotides) (Supplementary file 1 Table 5); insufficient data for fruit fly and yeast precluded the analysis for these species. The 5'UTR showed a ribosome profiling read density (translational efficiency, TE) comparable to that of the primary ORF in all datasets (Figure 4b). In contrast, the 3'UTR showed very little ribosome association both in codRNAs and lncRNAs. In fact, in many cases we could not find a single read mapping to the 3'UTR (31%-91% of cases in codRNAs and 46%-68% in lncRNAs). We also noted that the TE in the primary ORF of lncRNAs tended to be lower than the TE in the primary ORF of codRNAs, especially in human and mouse. Using genes with a single isoform or considering only annotated transcripts produced similar results (Supplementary file 1 Figures 7 and 8). We also controlled for expression level by dividing the dataset in transcripts with low (0.5-2 FPKM) and high expression (>2 FPKM), and by sampling the codRNAs in such a way as to have a similar expression distribution as lncRNAs in each of the two expression classes (Supplementary file 1 Figure 9). The results were very similar to those obtained with the complete dataset, indicating that the analysis is robust to transcript expression differences.

As lncRNAs generally contain more ORFs ≥ 24 amino acids than codRNAs (Supplementary file 1 Table 1), we performed a separate analysis in which we compared the translational efficiency of the primary ORF, any additional ORFs with mapped ribosome profiling reads, and the regions between ORFs protected by ribosomes (interORF) (Figure 4c, Supplementary file 1 Table 6). In species with sufficient data (the three vertebrate species and *Arabidopsis*), the TE of the primary ORF in codRNAs and lncRNAs was noticeably higher than the TE of the interORF regions (Wilcoxon rank-sum test, p-value $<10^{-9}$ in human, mouse, and zebrafish, p-value <0.05 in *Arabidopsis*). The data also indicated that ribosome binding is not always restricted to the primary ORF, especially in lncRNAs. Non-primary ORFs sequences from lncRNAs typically displayed higher TE values than the equivalent regions in codRNAs in human, mouse and zebrafish (Wilcoxon rank-sum test, p-value <0.001), consistent with the translation of several ORFs, some perhaps even shorter than our minimum size cutoff. Manual inspection of the ribosome profiling sequencing reads on lncRNA transcripts also supported this idea.

One example is AT1G34418.1, an annotated lncRNA in *Arabidopsis* that contains two instances of a 12 amino acid ORFs after the primary ORF, which are also covered by ribosome profiling reads (Supplementary file 1 Figure 2b). This case is reminiscent of the gene *pri* in fruit fly, which contains a monoexonic transcript translating several small redundant ORFs (Kondo *et al*, 2007).

Taken together, these results indicate that lncRNAs have ribosome profiling signatures consistent with translation, with a strong decrease of ribosome density in the 3'UTR but not the 5'UTR region, and preferential binding of ribosomes to the primary ORF. There exists the possibility that the translated peptides are degraded soon after being produced. However, we estimate that the percentage of cases that may undergo nonsense-mediated decay (NMD, see Methods for more details) is low, between 4.47 and 14.11% depending on the species. For comparison, the percentage for protein-coding transcripts showing the same patterns (including transcripts annotated as NMD in Ensembl) is between 0.34 and 13.33% (Supplementary file 1 Table 7).

**LncRNAs are lineage-specific**

We searched for homologues of primary ORFs in lncRNAs against codRNAs from the same species, (putative paralogues), or against primary ORFs in transcripts from the other species (putative orthologues) (Supplementary file 1 Table 8 and Supplementary file 2 Tables b and c, BLASTP with E-value $<10^{-4}$). The number of ORF in lncRNAs with homologous ORFs in other species was 0 in *Arabidopsis* and yeast, 1 in fruit fly (2.6%), 19 in mouse (5.2%), 27 in human (7.1%), and 124 in zebrafish (18.6%). In all cases the homologous ORFs corresponded to codRNAs. The number of lncRNAs with putative protein paralogues was 0 in *Arabidopsis*, yeast and fruit fly, 17 in mouse (4.6%), 22 in human (5.8%) and 161 in zebrafish (24.2%). These low numbers indicated that most lncRNAs are lineage-specific and do not represent non-functional duplicated gene copies. In the case of zebrafish, in which the values were substantially higher, there may be some conserved protein-coding genes that have not yet been annotated.

**Coding properties of ribosome-protected ORFs in lncRNAs**

Subsequently, we compared the sequence coding properties of the primary ORF in lncRNAs with those in *bona fide* coding and non-coding sequences. In each ORF, we computed a hexamer-based coding score based on the frequency of hexamers in coding sequences (codRNAs encoding experimentally validated proteins) *versus* the frequency in randomly taken intronic fragments (see Methods for more details). In fruit fly, the species with the lowest number of primary ORFs in lncRNAs, there were no significant differences between lncRNAs and intronic sequences. However, in the five other species the coding score of the primary ORF in lncRNAs (lncRNA_ribo), while lower than that of codRNAs, was significantly higher than the coding score of ORFs from introns (Figure 5, Wilcoxon rank-sum test, human, mouse, zebrafish and *Arabidopsis* p-value $<10^{-16}$, and yeast p-value $<10^{-5}$). This clearly indicates that lncRNAs are more coding-like than random ORFs. We repeated the same comparison using 100 different randomly sampled intronic sequence sets and in >95% of the cases we obtained the same result. LncRNAs associated with ribosomes (lncRNA_ribo) showed significantly higher coding scores than those not associated with ribosomes (lncRNA_noribo), even when we did not used the ribosome profiling information and compared the longest ORF in both types of transcripts (Supplementary file 1 Figure 12). We reached similar conclusions when we restricted the analysis to annotated lncRNA transcripts (Supplementary file 1 Figure 10) or when we used ORFs from gene deserts as an alternative non-coding sequence set (differences with lncRNAs significant by Wilcoxon rank-sum test, p-value $<10^{-16}$, see Methods for more details). Because a high proportion of lncRNAs contained small ORFs, we repeated the comparison only considering transcripts with ORFs shorter than 100 amino acids to avoid any length biases, again obtaining similar results (Supplementary file 1 Figure 11). The use of other coding scores based on codon frequencies instead of hexamers (dicodons), hexamer equiprobabilities instead of intronic hexamer, or sequence GC content, produced consistent results (Supplementary file 1 Table 9 and Figure 13).

At the individual transcript level, a sizeable fraction of lncRNAs associated with ribosomes displayed significantly higher coding scores than expected for non-coding sequences (Supplementary file 2 Table b, p-value <0.05 in all 100 intronic random sets). These transcripts are comprised of 177 human lncRNAs (43.7% of the lncRNAs, score >0.0189), 139 mouse lncRNAs (35.6%, score >0.0377), 567 zebrafish lncRNAs (78.1%, score >0.0095), 6 fruit fly lncRNAs (10.5%, score >-0.0483), 46 *Arabidopsis* lncRNAs (49.5%, score >-0.0202), and 22 yeast lncRNAs (49.9%, score >0.03387). Annotated and novel lncRNAs were present in similar proportions in these sets, supporting the validity of our strategy of merging the two types of transcripts from the beginning. Interestingly, quite a large fraction of these ORFs showed significant hits to predicted proteins in the NCBI non-redundant (nr) peptide database (22% to 57% depending on the species, Supplementary file 2 Table b).

If ORFs in lncRNAs are being translated this is likely to be a relatively recent evolutionary event, as many lncRNAs are lineage-specific (Pauli *et al*, 2012; Necsulea *et al*, 2014; our data). It is well established that proteins of different evolutionary age display distinct sequence properties, including different codon usage (Toll-Riera *et al*, 2009; Carvunis *et al*, 2012; Palmieri *et al*, 2014). We retrieved sets of annotated protein-coding transcripts of different evolutionary age from human, mouse, zebrafish, *Arabidopsis* and yeast, available from various studies (Ekman & Elofsson, 2010; Donoghue *et al*, 2011; Neme & Tautz, 2013), selecting only those that were expressed and associated with ribosomes in our study (Supplementary file 1 Table 10). We found that the coding score was always lower in the youngest group when compared to older groups (Figure 5, Wilcoxon rank-sum test, p-value <0.05). Remarkably, the coding scores of the youngest codRNAs were similar to those of lncRNAs (Figure 5). As an additional control, we also collected information from experimentally validated young proteins from the literature (Supplementary file 2 Table d). These proteins were short and the ORF occupied a relatively small fraction of the transcript. For example, in human the average size was 148 amino acids and transcript coverage 47%. The median coding score was remarkably low and again similar to that of lncRNAs (0.008 for primate-specific human transcripts, 0.046 for rodent-specific mouse transcripts, and 0.089 for yeast-specific coding transcripts). These results are consistent with the hypothesis that the majority of ribosome-protected ORFs in lncRNAs encode evolutionary young proteins.

**Selection pressure signatures in ORFs associated with ribosomes**

An important measure of the strength of purifying selection acting on a coding sequence is the ratio between the number of non-synonymous and synonymous single nucleotide polymorphisms (PN/PS). Given the nature of the genetic code, there are more possible non-synonymous mutations than synonymous mutations. Under neutrality (no purifying selection), the PN/PS ratio is expected to be approximately 2.89 (Nei & Gojobori, 1986).

Here we applied the large amount of available polymorphism data for human, mouse and zebrafish to compare the level of purifying selection in annotated codRNAs and in the primary ORFs in lncRNA sequences (Figure 6, Supplementary file 1 Table 11). In general, human sequences showed higher PN/PS ratios than sequences from the other species, probably due to the presence of many slightly deleterious mutations segregating in the population (Eyre-Walker, 2002). However, despite the differences between organisms, we observed the same general trends. First, the PN/PS was significantly lower in codRNAs than in lncRNAs (proportion test, p-value <$10^{-5}$), denoting stronger purifying selection in the former. Second, there was a very clear inverse relationship between the strength of purifying selection and the age of the gene (p-value $<10^{-15}$ between youngest and rest of codRNAs in mouse and zebrafish) as reported in previous studies (Liu *et al*, 2008; Cai *et al*, 2009). High PN/PS values were also observed in the subset of young genes encoding experimentally validated proteins in human (primate-specific median PN/PS of 3.10) and mouse (rodent-specific median PN/PS 1.42), confirming this tendency. Finally, the distribution of PN/PS values in lncRNAs was very similar to that of young protein-coding genes. In human and mouse there were no significant differences, and in the case of zebrafish the lncRNAs had even slightly lower PN/PS values than the youngest protein-coding gene class (p-value $<10^{-2}$). The latter result was in line with the high coding scores of lncRNAs in this species (previous section).

**DISCUSSION**

Here we analyzed the patterns of ribosome protection in polyA+ transcripts from cells belonging to six different eukaryotic species. Among the expressed transcripts we identified many lncRNAs in the different species. The vast majority of transcripts annotated as coding showed association with ribosomes (>92% in all species). Remarkably, a very large number of transcripts annotated as long non-coding RNA (lncRNAs) also showed such association (40-93% depending on the species). Considering that lncRNAs are typically much shorter and expressed at lower levels than codRNAs, which may hinder the identification of ribosome association, this is a very significant fraction. In addition the patterns of ribosome protection along the transcript are similar to those of protein-coding genes. Therefore, many lncRNAs appear to be scanned by ribosomes and translate short peptides.

Long non-coding RNAs are classified as such in databases because, according to a number of criteria, they are unlikely to encode functional proteins. These criteria includes the lack of a long ORF, the absence of amino acid sequence conservation and the lack of known protein domains (Harrow *et al*, 2012). Moreover, we expect lncRNAs not to have matches to proteomics databases, as this should classify them as coding. Annotated lncRNAs are typically longer than 200 nucleotides because this is the cutoff size normally implemented to differentiate them from other RNA classes such as microRNAs and small nuclear RNAs. In practice, it is difficult to classify a transcript as coding or non-coding on the basis of the ORF size (Dinger *et al*, 2008). Some true coding sequences may be quite small, and by chance alone non-coding transcripts may have relatively long ORFs. The majority of lncRNAs contain ORFs longer than 24 amino acids, which can potentially correspond to real proteins. Short proteins are more difficult to detect than longer ones and consequently they are probably underestimated in databases. In recent years, the use of comparative genomics (Frith *et al*, 2006; Ladoukakis *et al*, 2011; Hanada *et al*, 2013), proteomics (Slavoff *et al*, 2013; Vanderperre *et al*, 2013; Ma *et al*, 2014), and a combination of evolutionary conservation and ribosome profiling data (Crappé *et al*, 2013; Bazzini *et al*, 2014) have shown that the number of short proteins is probably much higher than previously suspected. In yeast, gene deletion experiments have provided evidence of functionality for short open reading frames (sORFs <100 amino acids) (Kastenmayer *et al*, 2006); in zebrafish, several newly discovered sORFs appear to be involved in embryonic development (Pauli *et al*, 2014) and other examples exist in human as well (Lee *et al*, 2013; Slavoff *et al*, 2014). In many cases, the transcripts containing sORFs will be classified as non-coding, especially if the ORF is not well conserved across different species.

One approach to identify potential coding transcripts is ribosome profiling (Ingolia *et al*, 2009), which has been used to study translation of proteins in a wide range of organisms (Guo *et al*, 2010; Ingolia *et al*, 2011; Brar *et al*, 2012; Michel *et al*, 2012; Chew *et al*, 2013; Dunn *et al*, 2013; Huang *et al*, 2013; Bazzini *et al*, 2014; Juntawong *et al*, 2014; Vasquez *et al*, 2014). In several of these studies it has been noted that lncRNAs can be protected by ribosomes (Ingolia *et al*, 2011; Chew *et al*, 2013; Bazzini *et al*, 2014; Juntawong *et al*, 2014). However, there is no consensus on whether the observed patterns are consistent with translation. For example in the original analysis of mouse stem cells, which we reanalyzed here, it was reported that many lncRNAs were polycistronic transcripts encoding short proteins (Ingolia *et al*, 2011), but in another paper where the same data was processed in a different way, they concluded that lncRNAs were unlikely to be protein-coding (Guttman *et al*, 2013). A zebrafish ribosome profiling study reported resemblance between lncRNAs and 5'leaders of coding RNAs; the authors suggested that translation may play a role in lncRNA regulation (Chew *et al*, 2013). Nevertheless, in the same study dozens of lncRNAs were proposed to be *bona fide* protein-coding transcripts. In *Arabidopsis*, the translational efficiency values of highly expressed lncRNAs (>5 FPKM) were similar to those of coding RNAs and some lncRNAs had profiles consistent with initiation and termination of translation (Juntawong *et al*, 2014). Finally, using yeast data, Wilson and Masel (2011) found many cases of non-coding transcripts bound to ribosomes and suggested that this facilitates the evolution of novel protein-coding genes from non-coding sequences.

The disparity of results obtained in different systems motivated us to retrieve the original data and perform exactly the same analyzes for six different species. As lncRNA catalogues are still very incomplete for most species, we also defined sets of novel lncRNAs using the RNA-seq sequencing reads for *de novo* transcript assembly. We discovered many novel, non-annotated, lncRNAs, especially in zebrafish, mouse, and fruit fly (Table 2). After the analysis of the ribosome profiling data, the same general picture emerged for the different biological systems, indicating that we are detecting very fundamental properties. In transcripts classified as lncRNAs, the ribosome profiling reads tend to cover a smaller fraction of the transcript than in typical codRNAs, in agreement with a shorter relative size of the ORF accumulating the largest number of ribosome profiling reads (primary ORF). We also find that the translational efficiency of regions corresponding to the primary ORF is much higher than that of 3'UTRs, both in codRNAs and lncRNAs, consistent with translation of the transcripts. This was confirmed by the manual inspection of the mapped ribosome profiling sequencing reads in the transcripts (examples in Supplementary file 1 Figure 2). Furthermore, with the exception of fruit fly for which we recovered relatively few lncRNAs associated with ribosomes, the primary ORF of lncRNAs showed significantly higher coding score than the longest ORF extracted from randomly selected non-coding regions.

LncRNAs are poorly conserved across species and so, if translated, they will produce species- or lineage-specific proteins. Recently evolved proteins are markedly different from widely distributed ancient proteins; they are shorter, subject to lower selective constraints and expressed at lower levels (Albà & Castresana, 2005; Levine *et al*, 2006; Cai *et al*, 2009; Liu *et al*, 2010; Xie *et al*, 2012; Donoghue *et al*, 2011). Here for the first time we have compared the properties of the ORFs in lncRNAs associated with ribosomes with the properties of annotated, and in some cases experimentally validated young protein-coding genes. LncRNAs and young protein-coding transcripts are virtually indistinguishable regarding their coding score and ORF selective constraints (Figure 5 and 6), strongly suggesting that many lncRNAs encode new peptides.

Although it is unclear how many of these peptides are functional, the data indicate that at least a fraction of them may be functional. Sequences that translate functional proteins are expected to display signs of selection related to preferential usage of certain amino acids and codons. This can be used to differentiate between coding and non-coding, especially in the absence of cross-species conservation, as is the case of most lncRNAs. About 35-40% of primary ORFs in human and mouse lncRNAs displayed coding scores that were significantly higher than those expected for non-coding sequences, making them excellent candidates for being functional. In fact, 1 mouse and 5 human lncRNAs associated with ribosomes that exhibited high coding scores in our study were re-annotated as protein-coding transcripts in a subsequent Ensembl gene annotation release (version 75, Supplementary file 2 Table c). Gene knock-out experiments in fly have discovered that young proteins, even if rapidly evolving, are often essential for the organism and can cause important defects when deleted (Chen *et al*, 2010; Reinhardt *et al*, 2013). Similarly, some peptides translated from lncRNAs may have important cellular functions yet to be discovered.

LncRNAs are expressed at much lower levels than typical protein-coding genes, and this means that the translated peptides would also be produced in small quantities. It may be that some of these peptides are not functional, but their translation does not produce a large enough deleterious effect for them to be eliminated via selection. Pseudogenes also showed extensive association with ribosomes in our study, indicating that the translation machinery is probably not very selective. The data also indicate that a fraction of lncRNAs have not acquired the capacity to be translated. Depending on the experiment analyzed, a variable fraction of lncRNAs did not show any significant association with ribosomes. As previously discussed, this is probably affected by a lack of sensitivity; it is also true that the lncRNAs not associated with ribosomes tended to show lower coding scores than lncRNAs associated with ribosomes, even when we did not use the ribosome profiling data and simply compared the longest ORF in both kinds of transcripts.

Recently, it has been reported that human-specific protein-coding genes are often related to non-coding transcripts in macaque, pointing to a non-coding origin for many newly evolved proteins (Xie *et al*, 2012). More generally, one may view *de novo* protein-coding gene evolution as a continuum from non-functional genomic sequences to fully-fledged protein-coding genes (Levine *et al*, 2006; Toll-Riera *et al*, 2009; Carvunis *et al*, 2012). Therefore, many lncRNAs could be intermediate states in this process, their pervasive translation serving as the building material for the evolution of new proteins. It may be difficult to obtain functional proteins from completely random ORFs (Jacob, 1977), but the effect of natural selection preventing the production of toxic peptides (Wilson & Masel, 2011), and the high number of transcripts expressed in the genome, may facilitate this process.

**METHODS**

***Sequencing and mapping of reads***

We downloaded the original data from Gene Expression Omnibus (GEO) for six different ribosome profiling experiments that had both ribosome footprinting and polyA+ RNA-seq sequencing reads: mouse (*Mus musculus*) (Ingolia *et al*, 2011), human (*Homo sapiens*) (Guo *et al*, 2010), yeast (*Saccharomyces cerevisiae*) (Brar *et al*, 2012), zebrafish (*Danio rerio*) (Chew *et al*, 2013), fruit fly (*Drosophila melanogaster*) (Dunn *et al*, 2013) and *Arabidopsis* (*A. thaliana*) (Juntawong *et al*, 2014). We retrieved genome sequences and gene annotations from Ensembl v.70 and Ensembl Plants v.21 (Flicek *et al*, 2012).

Raw ribosome and RNA-seq sequencing reads underwent quality filtering using Condentri (v.2.2) (Ku, 2011) with the following settings (-hq=30 –lq=10). Adaptors described in the original publications were trimmed from filtered reads if at least 5 nucleotides of the adaptor sequence matched the end of each read. In zebrafish, reads from different developmental stages were pooled to improve read coverage. In all experiments, reads below 25 nucleotides were not considered. Clean ribosome short reads were filtered by mapping them to the corresponding species reference RNA (rRNA) using the Bowtie2 short-read alignment program (v. 2.1.0) (Langmead *et al*, 2009). Unaligned reads were aligned to a genomic reference genome with Bowtie2 allowing one mismatch in the first 'seed' region (the length of this region was selected according to the descriptions provided in each individual experiment). RNA-seq short reads were mapped with Tophat (v. 2.0.8) (Kim *et al*, 2013) to the corresponding reference genome. We allowed two mismatches in the alignment with the exception of zebrafish, for which we allowed three mismatches since the reads were significantly longer. Multiple mapping was allowed unless specifically stated.

*Defining a set of expressed transcripts*

Expressed transcripts were assembled using Cufflinks (v 2.2.0) (Trapnell *et al*, 2010). We initially considered a transcript as expressed if it was covered by at least 4 reads and its abundance was higher than 1% of the most abundant isoform of the gene. We also discarded assembled transcripts in which >20% of reads were mapped to several locations in the genome. Gene annotation files from Ensembl (gtf format, v.70) were provided to Cufflinks to guide the reconstruction of already annotated transcripts. Annotated transcripts were divided into coding RNAs and long non-coding RNAs (lncRNAs), we only considered lncRNAs that were not part of genes with coding transcripts. Novel isoforms corresponding to annotated loci were not analyzed. Transcripts that did not match or overlapped annotated genes were labeled 'novel' lncRNAs. We used a length threshold of 200 nucleotides to select novel long non-coding RNAs, as in ENCODE annotations (Djebali *et al*, 2012).

Strand directionality of multiexonic transcripts was inferred using the splice site consensus sequence. We only considered monoexonic transcripts in the case of fruit fly, *Arabidopsis* and yeast, provided the transcripts were intergenic. This was motivated by the observation that, in human, mouse, and zebrafish, only about 2% of the annotated coding genes were single exon genes, whereas this value was 9.3% and 13.9% in *Arabidopsis* and fruit fly, respectively. In addition, many of the recently evolved genes in these species are monoexonic (Ekman & Elofsson, 2010; Donoghue *et al*, 2011).

The inclusion of novel lncRNAs made it possible to perform analyzes of species for which there are very few annotated lncRNAs. In yeast, we only found 3 annotated lncRNA but we could reconstruct 69 novel ones. As UTR regions in yeast genes are not annotated in Ensembl because of the variability seen in transcription start sites (TSS), we downloaded all defined yeast 5' UTR (Miura *et al*, 2006) to see whether some lncRNAs could be part of UTRs. Only 4 of them were good candidates to be UTR regions from annotated transcripts, 3 of them exhibiting ORFs associated with ribosomes. These cases have been indicated in Supplementary file 2.

Coding transcripts were classified into different subclasses depending on the existing annotations: a) Annotated protein-coding transcripts (codRNA). b) Annotated transcripts with surveillance mechanisms (nonsense mediated decay, nonstop mediated decay and no-go decay). c) Annotated pseudogenes. We removed protein-coding transcripts in which annotated coding sequences (CDS) are still incomplete.

Subsequently, we defined an additional subset of annotated protein-coding transcripts with well-established coding properties based on the existence of an experimentally verified protein in Swiss-Prot for each gene ("evidence at protein level", downloaded 29th October 2013, UniProt Consortium, 2014). These transcripts were labeled codRNAe. The rest of annotated protein-coding transcripts were abbreviated codRNAne. In zebrafish, most proteins are not yet experimentally validated and therefore we joined all protein-coding transcripts into a single group.

We built a dataset of human lncRNAs with described non-coding functions using data obtained from several recent reviews (Ponting *et al*, 2009; Ulitsky & Bartel, 2013; Fatica & Bozzoni, 2014). This dataset included 29 different genes (Supplementary file 2 Table a).

We used cufflinks to estimate the expression level of a transcript in FPKM units (Fragments Per Kilobase per total Million mapped reads). For transcripts encoding experimentally validated proteins and expressed at <0.5 FPKM detection of ribosome binding was not optimal, especially in zebrafish and human, in which only ~50% and ~85% of the transcripts showed association with ribosomes. Using a threshold of >0.5 FPKM, we ensured ~75% of zebrafish and ~96% of human protein-coding transcripts were associated with ribosomes, whereas in the rest of species ribosome association was detected in nearly 100% of cases. Consequently, we employed >0.5 FPKM as the expression threshold for transcript analysis.

*Definition of potential open reading frames (ORFs) and other transcript regions*

We predicted all possible open reading frames (ORFs) in the expressed transcripts. We defined an ORF as any sequence starting with an AUG codon and finishing at a stop codon (TAA, TAG or TGA), and at least 75 nucleotides long. This would correspond to a 24 amino acid protein, which is the size of the smallest complete human polypeptide found in genetic screen studies (Hashimoto *et al*, 2001). This ORF definition will not detect non-canonical ORFs with different start or stop codons, although these ORFs often correspond to regulatory ORFs (uORFs) in the 5' UTR region. In monoexonic transcripts we considered all 6 possible different frames.

We also defined each transcript 5' UTR as the region between the transcription start site and the AUG codon from the left-most predicted ORF, and the 3' UTR the region from the stop codon in the right-most predicted ORF to the transcription end site. UTRs with lengths below 30 nucleotides were not analyzed since ribosome reads could not be properly aligned to these regions due to their small size. Regions between two consecutive putatively translated ORFs (with ribosome profiling reads) were termed interORFs. Only interORFs of size 30 nucleotides or longer were analyzed.

Next, we defined a set of *bona fide* non-coding sequences sampled from intronic fragments. We used the introns of the genes expressed in each experiment, provided they did not overlap to any exons from other overlapping genes. We randomly selected fragments in such a way as to simulate the same size distribution as in the complete set of expressed transcripts. We performed 100 simulations of intron sampling to ensure the results were robust to the randomization process. We selected the longest ORF in each intronic fragment for the calculation of coding scores and GC content (see below).

*Association with ribosomes and translational efficiency (TE)*

We computed the number of reads overlapping each feature of interest (transcript, UTR, ORF and interORF) using the BEDTools package (v. 2.16.2) (Quinlan & Hall, 2010). We only considered ribosome reads in which more than half of their length spanned the considered region, since the ribosome P-site is usually detected at the central region of the read, with small differences depending on each experiment, length of the read and species. We set up a minimum ribosome profiling coverage of 75 nucleotides per transcript to define the transcript or transcript region (e.g. ORF) as associated with ribosomes. This is significantly longer than the length of the ribosome profiling sequencing reads (36 to 51 nucleotides) and is consistent with the minimum ORF length threshold.

The translational efficiency (TE) of a sequence has been previously defined as the density of ribosome profiling (RPF) reads normalized by transcript abundance (Ingolia *et al*, 2009). We calculated it by dividing the FPKM of the ribosome profiling experiment by the FPKM of the RNA-seq experiment.

In the case of transcripts we divided the sequence in 90 nucleotide windows and obtained two different scores: maximum TE score in a window (maxTE) and mean TE score between regions covered by ribosome reads in at least 75 nucleotides (meanTE). We employed these measures because the ribosomes are not expected to cover the complete transcript.

We defined the primary ORF in a transcript associated with ribosomes as the ORF with the largest number of RPF reads with respect to the total RPF reads covering the transcript. The rest of ORFs associated with ribosomes were considered as well; in the cases where two or more ORFs overlapped, we selected the longest one. In ORFs, interORFs and UTRs, we computed the TE along the whole region. We only compared the TE values in different regions in transcripts in which all the compared regions showed >0.2 FPKM.

*Peptide evidence in existing proteomics databases*

We downloaded all peptide sequences from the PeptideAtlas database: 338,013 human peptides (August 2013), 101,695 mouse peptides (June 2013), 86,836 yeast peptides (March 2013), and 58,746 fruit fly peptides (August 2012). We investigated if the number of ribosome-associated protein-coding transcripts that matched the peptides in these databases varied with protein length. We omitted this analysis in zebrafish and *Arabidopsis* due to the lack of sufficiently large peptide databases. The matches were identified using BlastP searches (Altschul *et al*, 1997). We selected perfect matches only.

*Evidence of nonsense mediated decay in ORFs*

We investigated how many of these predicted ORFs may be candidates for being regulated via non-sense mediated decay (NMD) surveillance pathways, whose main function is to eliminate transcripts containing premature stop codons. We defined NMD candidates as all cases in which the stop-codon from a predicted ORF was located ≥ 55 nucleotides upstream of a splice junction site, provided the stop-codon was not in the terminal exon (Scofield *et al*, 2007). Although this mechanism is well characterized in protein-coding genes, it was previously proposed that it may be also a way to degrade non-functional peptides translated in lncRNAs (Tani *et al*, 2013). Other surveillance mechanisms, such as non-stop mediated decay or no-go decay, were not considered since all predicted ORFs finished at a stop codon and we did not analyze RNA secondary structures.

*Defining ages of protein-coding transcripts*

We utilized existing gene age classifications in human, mouse and zebrafish (Neme & Tautz, 2013) to identify young gene classes: human primate-specific (~55.8 My), mouse rodent-specific (~61.7 My), human and mouse mammalian-specific (~225 My), zebrafish actinopterygii-specific (~420 My) (abbreviated fish) and metazoan (~800 My). In yeast we used predefined genes specific to *Saccharomyces cerevisiae* (1-3 My)(abbreviated S. cerevisiae) and the *Saccharomyces* group (~100 My) (Ekman *et al*, 2007). In *Arabidopsis* we retrieved Cruciferae(Brassicaceae)-specific genes (20-40 My) (Donoghue *et al*, 2011). These genes are believed to have arisen primarily by *de novo* mechanisms, as no homologies in other species have been detected despite the fact that many closely related genomes have now been sequenced.

*Defining gene desert sequences*

In human, we added a set of gene desert sequences as defined in Ovcharenko et al. (2005). We selected two stable and two flexible gene deserts, depending on the rate of conservation in other species; both are from chromosomes 4 (flexible located in coordinates 136,000,001-138,000,000; stable located in coordinates 180,000,001-182,000,010), that is comprised of a high number of gene deserts, and 17

(flexible located in coordinates 51,100,001-51,900,000; stable located in coordinates 69,300,001-70,000,000), which has a higher gene density. We ensured that no protein-coding genes were annotated in later Ensembl versions in these regions. We predicted all possible ORFs in these regions to evaluate coding scores and GC content.

*Evaluation of codon and dicodon (hexamer) usage score in ORFs*

The examination of nucleotide hexamer frequencies has been shown to be a powerful way to distinguish between coding and non-coding sequences (Sun *et al*, 2013; Wang *et al*, 2013). We initially computed one coding score (CS) per nucleotide hexamer:

$$CS_{hexamer(i)} = \log\left(freq_{coding}(hexamer(i)) \big/ freq_{non-coding}(hexamer(i))\right)$$

The coding hexamer frequencies were obtained from all annotated coding sequences in protein-coding transcripts encoding experimentally validated proteins (except for zebrafish in which all protein-coding transcripts were considered). The non-coding hexamer frequencies were calculated using the longest ORF in intronic regions that were selected randomly from expressed protein-coding genes. Next, we used the following statistic to measure the coding potential of a sequence:

$$CS_{ORF} = \frac{\sum_{i=1}^{i=n} CS_{hexamer(i)}}{n}$$

where *i* is each sequence hexamer in the ORF, and *n* the number of hexamers considered.

The hexamers were calculated in steps of 3 nucleotides in frame (dicodons). We did not consider the initial hexamers containing a Methionine or the last hexamers containing a stop codon, since they are not informative. Given that all ORFs were at least 75 nucleotides long the minimum value for *n* was 22.

We calculated other related statistics in a similar way. This included using an equiprobable hexamer distribution instead of the distribution obtained from non-coding sequences, or using codon frequencies instead of hexamer frequencies. These statistics showed somewhat lower power to distinguish between coding and non-coding sequences. As a complementary measure, we quantified the GC content in different coding and non-coding transcripts and ORFs.

*Sequence similarity searches*

We employed BLASTP with an E-value cutoff of $10^{-4}$ to compare the amino acid sequences encoded by ORFs in different kinds of transcripts. We searched for lncRNA homologues in the NCBI non-redundant (nr) protein database (Pruitt *et al*, 2014) using BLASTX (Evalue $<10^{-4}$). BLAST sequence similarity search programs are based on gapped local alignments (Altschul *et al*, 1997).

*Analysis of genomic polymorphisms*

We downloaded all available single-nucleotide polymorphisms (SNPs) from dbSNP (Sherry *et al*, 2001) for human (~50 million), mouse (~64.2 million), and zebrafish (~1.3 million). We classified SNPs in ORFs as non-synonymous (PN, amino acid altering) and synonymous (PS, not amino acid altering). We computed the PN/PS ratio in each sequence dataset by using the sum of PN and PS in all sequences. The estimation of PN/PS ratios of individual sequences was in general not reliable due to lack of sufficient SNP data. We obtained confidence intervals using the proportion test in R (see below).

*Statistical data analyzes*

The analysis of the data, including generation of plots and statistical tests, was done with R (R Development Core Team, 2010).

**SUPPLEMENTARY FILES**

Supplementary files can be found at our web portal http://evolutionarygenomics.imim.es. Supplementary file 1 is available at http://evolutionarygenomics.imim.es/group/Ruiz-Orera_etal_sup1.pdf and contains additional figures and tables mentioned in the text. Supplementary file 2 is available at http://evolutionarygenomics.imim.es/group/Ruiz-Orera_etal_sup2.xls and contains gene and transcript datasets.

**ACKNOWLEDGEMENTS**

We acknowledge Jose Luis Villanueva Cañas and Will Blevins for their critical revision of the manuscript. We are grateful to Ivan Ovcharenko for advise on gene deserts. This work was funded by Ministerio de Economía y Competitividad (FPI to J.R-O., BFU2012-36820) and Fundació ICREA (MM.A.).

**FIGURE LEGENDS**

**Figure 1. Comparison between coding RNA and lncRNA transcripts.** A) Density plots of transcript length. B) Box-plots of transcript expression level in log2(FPKM) units. lncRNA_ribo: lncRNAs associated with ribosomes; lncRNA_noribo: lncRNAs for which association with ribosomes was not detected. codRNA: coding transcripts encoding experimentally validated proteins except for zebrafish in which all transcripts annotated as coding were considered. The area within the box-plot comprises 50% of the data and the line represents the median value. In all studied species, codRNAs had significantly stronger expression values than lncRNAs (Wilcoxon rank-sum test, p-value $<10^{-5}$), and, except for yeast, lncRNA_ribo were also significantly more expressed than lncRNA_noribo in terms of FPKM (Wilcoxon rank-sum test, p-value <0.005).

**Figure 2. Effect of transcript expression level on the detection of ribosome association.** The percentage of transcripts associated with ribosomes is shown for several transcript expression intervals. codRNA: annotated coding transcripts encoding experimentally verified proteins (except in zebrafish for which all coding transcripts were considered). lncRNA: annotated and novel long non-coding RNAs. Only species with at least 20 transcripts in each expression bin were plotted. In the rest of species the data was consistent with the trends shown.

**Figure 3. Ribosome association profiles for coding and lncRNA transcripts.** Box-plots of transcript mean translational efficiency (TE) in log2(TE) units. The area within the box-plot comprises 50% of the data and the line represents the median value. lncRNA: lncRNAs for which association with ribosomes was detected. codRNA: coding RNAs transcripts encoding experimentally validated proteins except for zebrafish in which all transcripts annotated as coding were considered.

**Figure 4. Ribosome association in different transcript regions.** A) Density plot of the relative length of the primary ORF with respect to transcript length. B) Box-plots of TE distribution in primary ORF, 5'UTR and 3'UTR regions. The area within the box-plot comprises 50% of the data and the line represents the median value. The analysis considered all transcripts with 5'UTR and 3'UTR longer than 30 nucleotides and > 0.2 FPKM in all three regions. C) Box-plots of TE distribution in primary ORFs, rest of ORFs with ribosome profiling reads and non-ORF regions (interORF). The analysis considered all transcripts with at least two ORFs and more than 30 nucleotides interORF. Fruit fly and yeast were not included in this analysis due to insufficient data.

**Figure 5. Coding scores in primary ORFs from different types of transcripts.** The coding score was calculated as the log ratio of hexamer frequencies in coding versus intronic sequences. In lncRNA_noribo and introns, the longest ORF was considered. The Class 'pseudogene' was only included in species with more than 20 expressed pseudogenes with mapped ribosome profiling reads. The coding score of the primary ORF in lncRNAs (lncRNA_ribo) was significantly higher than the coding score in ORFs defined in introns (Wilcoxon rank-sum test, human, mouse, zebrafish and *Arabidopsis* p-value $<10^{-16}$, and yeast p-value $<10^{-5}$, Wilcoxon rank-sum test), in lncRNA_ribo higher than lncRNA_noribo (Wilcoxon rank-sum test, human, mouse and zebrafish p-value $<10^{-5}$, and *Arabidopsis* p-value $<0.05$). Young codRNAs displayed similar scores than lncRNA_ribo in mouse, zebrafish and yeast, being only significantly higher in human and *Arabidopsis* (Wilcoxon rank-sum test, p-value $<0.005$).

**Figure 6. Selective pressure in ORFs from different types of transcripts.** PN/PS: ratio between the number of non-synonymous and synonymous single nucleotide polymorphisms in the complete set of primary ORFs for a given class of transcripts (in lncRNA_noribo the longest ORF was considered). In blue, data for different coding and non-coding transcript classes. In brown, data for different age codRNA classes. The bars represent the 95% confidence interval for the PN/PS value.

**TABLES**

**Table 1. Datasets used in the study.**

| Species | | GEO Accession | Mapped reads (millions) | Max Read Length (bp) | Description | Reference |
|---|---|---|---|---|---|---|
| Mouse *M.musculus* | RNA-seq | GSE30839 | 226.0 | 43 | ES cells, E14 | Ingolia et al., 2011 |
| | Ribosome Profiling | GSE30839 | 39.2 | 47 | | |
| Human *H.sapiens* | RNA-seq | GSE22004 | 29.8 | 36 | HeLa cells | Guo et al., 2010 |
| | Ribosome Profiling | GSE22004 | 78.3 | 36 | | |
| Zebrafish *D.rerio* | RNA-seq | GSE32900 | 1382.2 | 2x75 | Series of developmental stages | Chew et al., 2013 |
| | Ribosome Profiling | GSE46512 | 1040.0 | 44 | | |
| Fruit fly *D.melanogaster* | RNA-seq | GSE49197 | 1317.9 | 50 | 0-2h embryos, wild-type | Dunn et al., 2013 |
| | Ribosome Profiling | GSE49197 | 105.7 | 50 | | |
| Arabidopsis *A.thaliana* | RNA-seq | GSE50597 | 79.8 | 51 | No stress conditions, TRAP purification | Juntawong et al., 2014 |
| | Ribosome Profiling | GSE50597 | 140.3 | 51 | | |
| Yeast *S.cerevisiae* | RNA-seq | GSE34082 | 3.57 | 36 | Starved conditions, strain SK1 | Ingolia et al., 2009 |
| | Ribosome Profiling | GSE34082 | 6.56 | 36 | | |

**Table 2. Fraction of transcripts associated with ribosomes.**

|  | codRNA | annotated lncRNA | novel lncRNA |
|---|---|---|---|
| **Mouse** | 14,196/14,245 (99.7%) | 218/253 (86.2%) | 172/223 (77.1%) |
| **Human** | 16,630/17,011 (97.8%) | 403/846 (47.4%) | 2/86 (2.3%) |
| **Zebrafish** | 11,643/12,595 (92.4%) | 130/322 (40.4%) | 596/2070 (28.8%) |
| **Fruit fly** | 8,031/8,041 (99.9%) | 15/16 (93.7%) | 42/108 (38.9%) |
| **Arabidopsis** | 18,879/19,162 (98.5%) | 55/61 (90.2%) | 38/78 (48.7%) |
| **Yeast** | 5,109/5,473 (93.3%) | 2/3 (66.7%) | 47/69 (68.1%) |

**Table 3. Fraction of translated proteins of different size detected in proteomics databases.** Only transcripts encoding experimentally validated proteins (codRNAe) were considered.

|  | Protein size (amino acids) | | | |
| --- | --- | --- | --- | --- |
| Species | 24-80 | 81-130 | 131-180 | >180 |
| Mouse | 27 / 58 (46.6%) | 222 / 286 (77.6%) | 256 / 330 (77.6%) | 3716 / 4786 (77.7%) |
| Human | 116 / 272 (42.6%) | 536 / 748 (71.7%) | 669 / 875 (76.5%) | 6757 / 8964 (75.4%) |
| Yeast | 33 / 42 (78.6%) | 186 / 255 (72.9%) | 261 / 302 (86.4%) | 3575 / 4024 (88.8%) |

**FIGURES**

**Figure 1**

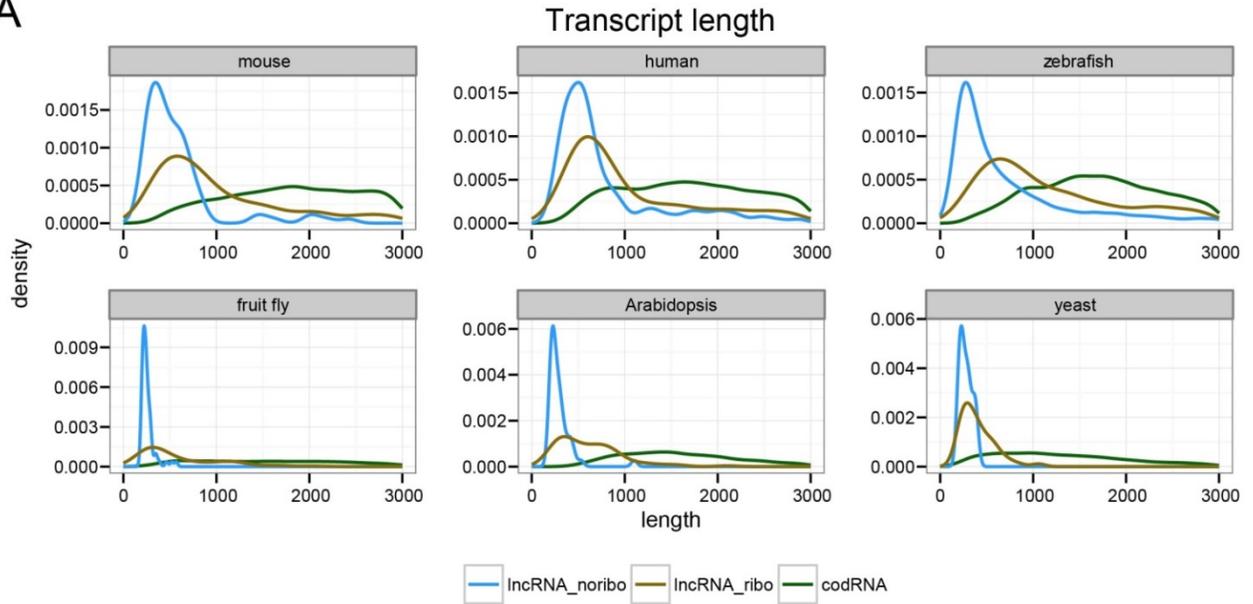

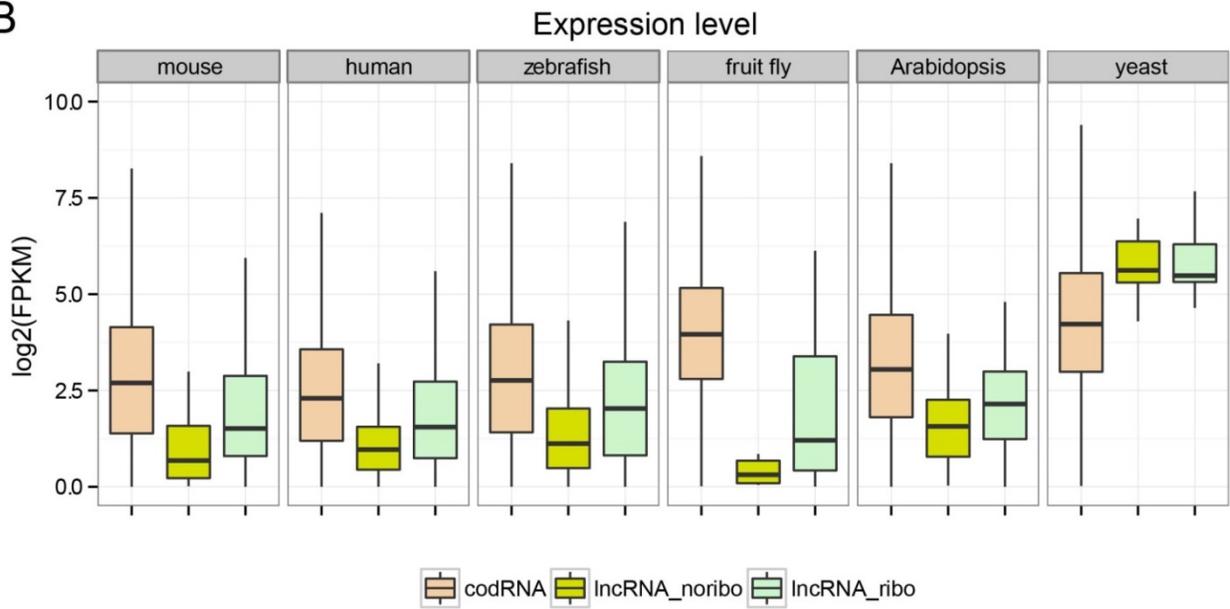

**Figure 2**

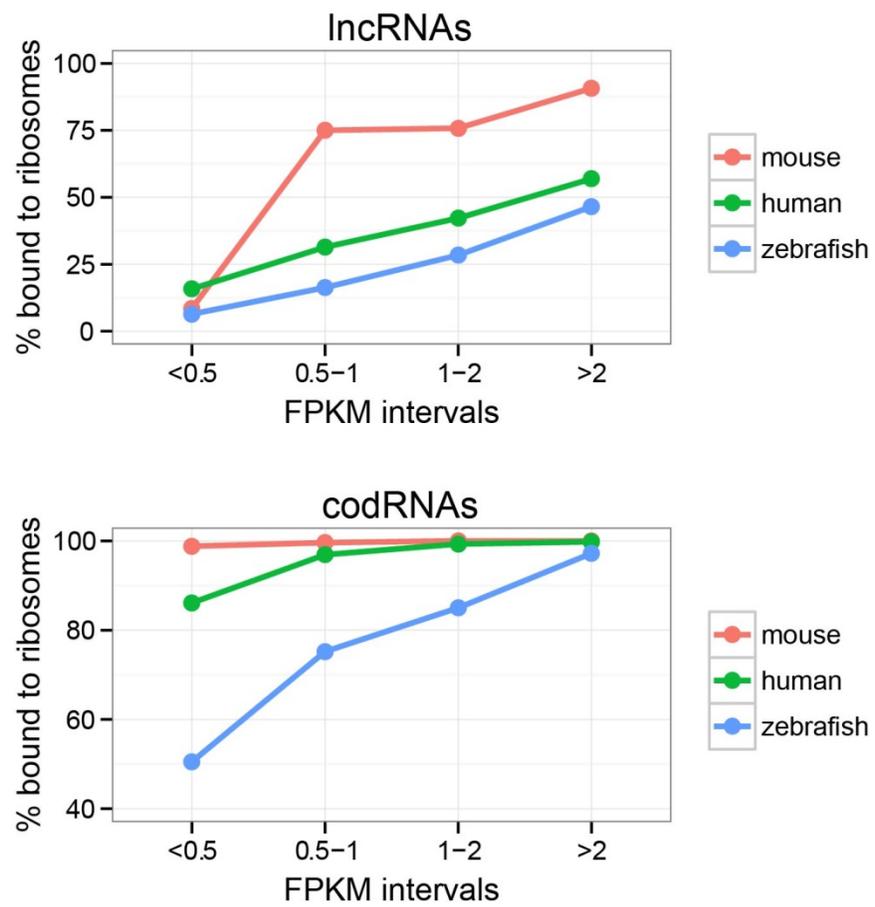

**Figure 3**

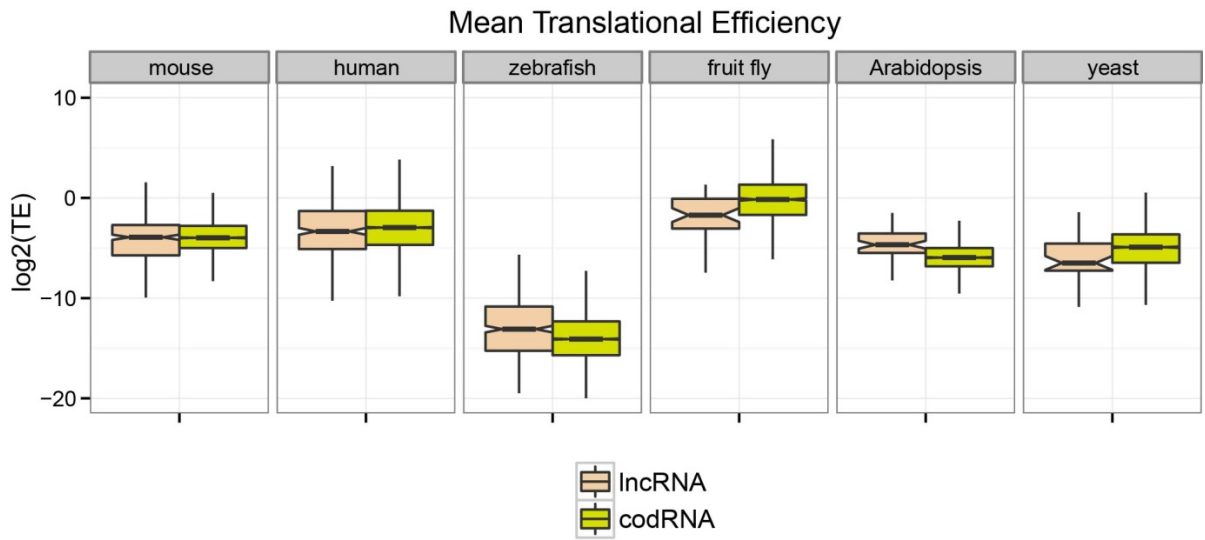

**Figure 4**

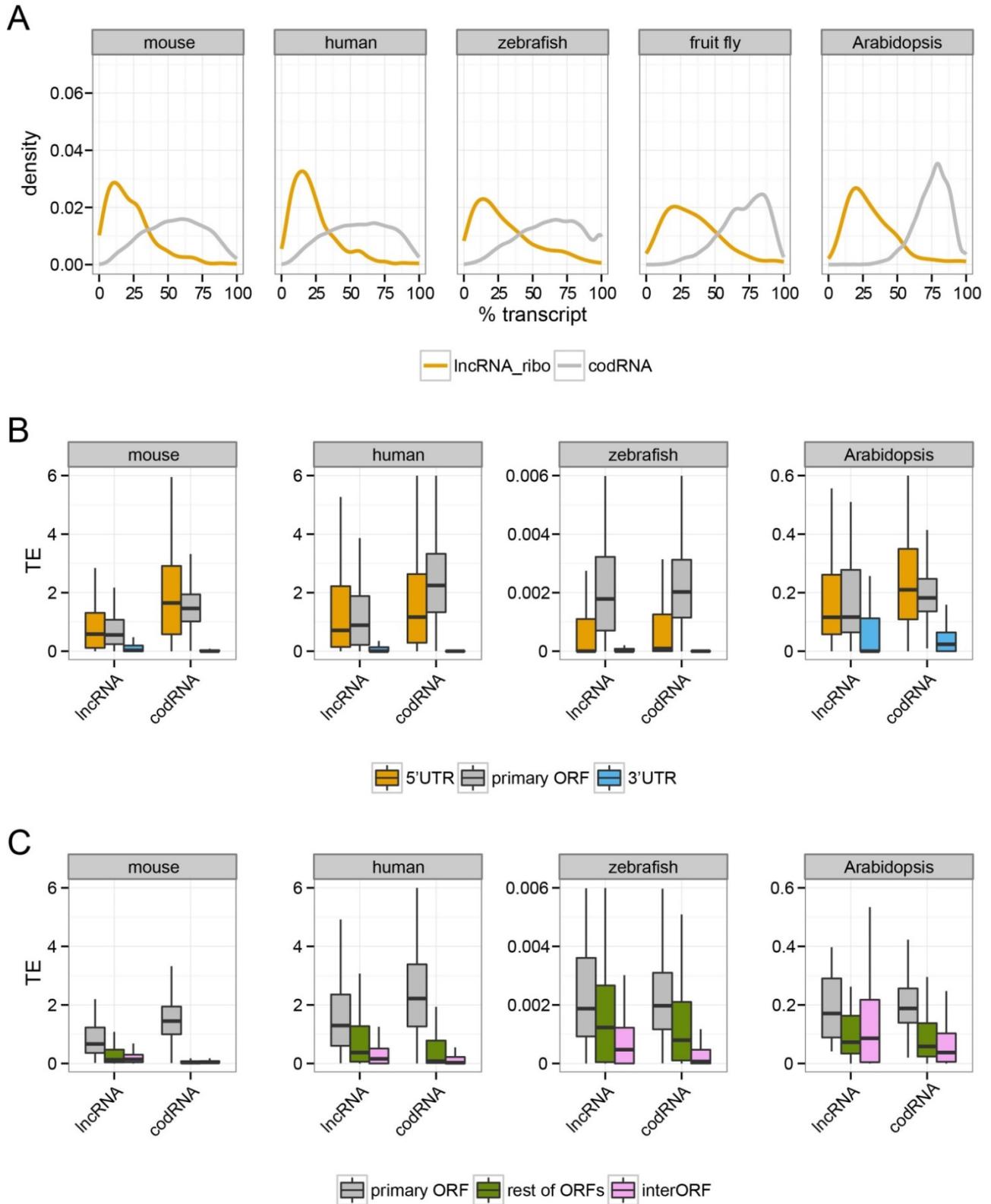

**Figure 5**

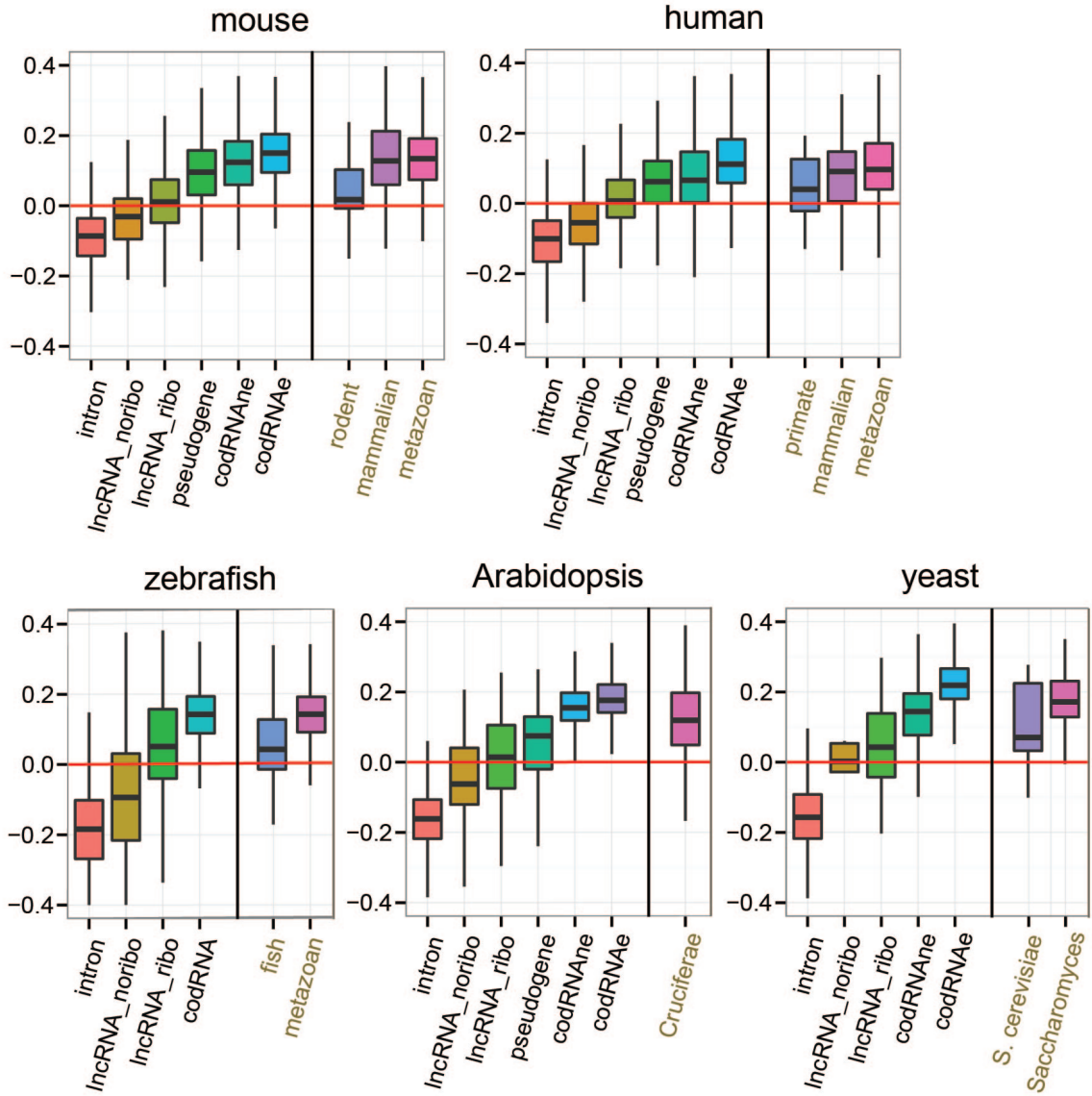

**Figure 6**

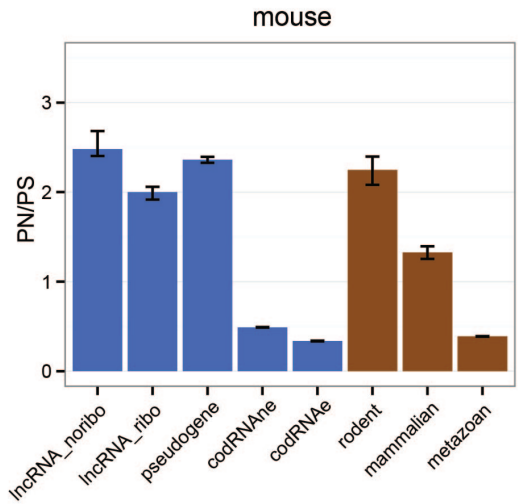

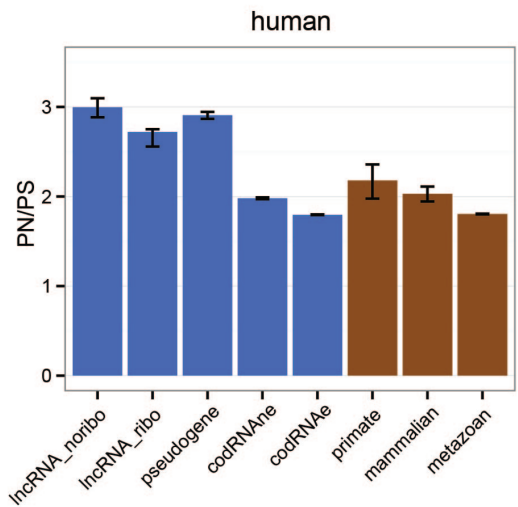

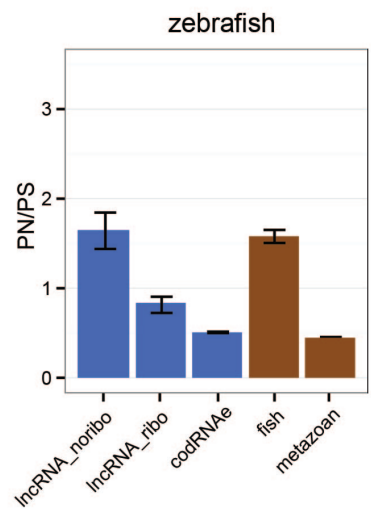